\pdfoutput=1
\documentclass[prd,
               twocolumn,
               amssymb,
               amsmath,
               eqsecnum,
               showpacs,
               letterpaper,
               superscriptaddress,
               altaffilletter]
               {revtex4}

\usepackage{color}
\usepackage{graphicx}
\usepackage{latexsym}
\usepackage{float}
\usepackage{amsmath}
\usepackage{amssymb}
\usepackage{multirow}

\begin{document}

\title{Localization of gravitational wave sources with networks of advanced detectors}

\author{S.~Klimenko}  
\affiliation{University of Florida, P.O.Box 118440, Gainesville, Florida, 32611, USA}
\author{G.~Vedovato}  
\affiliation{INFN, Sezione di Padova, via Marzolo 8, 35131 Padova, Italy}
\author{M.~Drago}  
\affiliation{University of Trento, Physics Department and INFN, Gruppo Collegato di Trento, 
             via Sommarive 14, 38123 Povo, Trento, Italy}
\author{G.~Mazzolo} 
\affiliation{Max Planck Institut f\"{u}r Gravitationsphysik, Callinstrasse 38, 
             30167 Hannover, and Leibniz Universit\"{a}t Hannover, Hannover, Germany}
\author{G.~Mitselmakher}
\affiliation{University of Florida, P.O.Box 118440, Gainesville, Florida, 32611, USA}
\author{C.~Pankow} 
\affiliation{University of Florida, P.O.Box 118440, Gainesville, Florida, 32611, USA}
\author{G.~Prodi} 
\affiliation{University of Trento, Physics Department and INFN, Gruppo Collegato di Trento, 
             via Sommarive 14, 38123 Povo, Trento, Italy}
\author{V.~Re} 
\affiliation{INFN, Sezione di Roma Tor Vergata, Via della Ricerca Scientifica 1, 00133 Roma, Italy}
\author{F.~Salemi} 
\affiliation{Max Planck Institut f\"{u}r Gravitationsphysik, Callinstrasse 38, 
             30167 Hannover, and Leibniz Universit\"{a}t Hannover, Hannover, Germany}
\author{I.~Yakushin} 
\affiliation{LIGO Livingston Observatory, LA , USA}

\begin{abstract}
Coincident observations with gravitational wave (GW) detectors and other astronomical
instruments are in the focus of the experiments with the network of 
LIGO, Virgo and GEO detectors. They will become a necessary part of the future GW astronomy 
as the next generation of advanced detectors comes online.
The success of such joint observations directly depends on the source localization capabilities
of the GW detectors. In this paper we present studies of the sky localization of
transient GW sources with the future advanced detector networks and describe their fundamental properties. 
By reconstructing sky coordinates of ad hoc signals injected into simulated detector noise 
we study the accuracy of the source localization and its dependence on the strength of 
injected signals, waveforms and network configurations.

\end{abstract}

\date[\relax]{Dated: \today }

\maketitle

\section{Introduction}

There has been a significant sensitivity improvement of the
gravitational wave detectors since the Laser Interferometer Gravitational Wave Observatory 
(LIGO)~\cite{LIGO} and Virgo observatory~\cite{Virgo} started their operation.
In 2007 LIGO and Virgo completed the two year run at sensitivity
that allows detection of a merger of two neutron stars (NS-NS) as far as $\sim{30}$~Mpc 
away~\cite{S5range, cbc2009}.
In the most recent run (May 2009 - October 2010) the binary neutron star horizon 
distance has been increased to $\sim{40}$~Mpc. However, even at this impressive sensitivity,
the anticipated detection rate with the initial LIGO and Virgo detectors is quite 
low. A detection may be possible in the case of a rare astrophysical transient
event such as
a supernova explosion in our Galaxy or a nearby merger of binary neutron stars. 
The signal is likely to be weak and it will be difficult to prove its astrophysical 
origin unless it is confirmed with a coincident observation of the 
electromagnetic or neutrino counterpart. For this reason the LIGO and Virgo collaborations
are conducting a wide range of joint observations~\cite{LOOKUP} 
with other astrophysical experiments including radio~\cite{LOFAR, ARECIBO}, 
optical and x-ray telescopes~\cite{ROTSE, QUEST, TAROT, SWIFT}, and neutrino 
detectors~\cite{Antares, IceCube}. 

A more robust detection of gravitational waves from
astrophysical sources is anticipated in the next five years as Advanced LIGO
and Advanced Virgo come online. Numerous GW signals, expected to be observed by advanced
detectors (likely $\sim{40}$ NS-NS events per year~\cite{ExpRates}), 
will begin our exploration of the gravitational-wave sky and start the era of 
the gravitational wave astronomy.  Along with the advanced GW detectors, a new
generation of optical telescopes will come online
~\cite{LSST, PanStarrs, 30mTelescope}, which
will  enable a wide and deep survey of the electromagnetic sky.
Joint observations with the advanced  gravitational
wave detectors and electromagnetic instruments will not only 
increase the confidence of detection but also bring fundamentally new
information about the GW sources. They will reveal
the physics and dynamic of sources, provide the identification of host
galaxies and the associated redshifts, and in some cases determine luminosity 
distance to the source. 

One of the major challenges for such joint observations is to establish unambiguous 
association between a gravitational wave signal and a possible electromagnetic counterpart. 
It greatly depends on the ability of the GW networks to reconstruct sky coordinates
of a detected GW source. Given an accurate sky location, a corresponding electromagnetic 
transient may be identified in a list of events obtained with the all-sky telescope
surveys, or the EM instruments can be guided to take images of a small area in the sky.
In the second case, it is important that the sky localization  is performed 
by GW detectors in real time with low latency.
The efficiency of the GW-EM association and the choice of a partner telescope is affected 
by the sky localization error which should be well within the instrument's field of view
(typically less than few square degrees).
Moreover, exploring smaller area in the sky will decrease the probability of 
the false association. 

The problem of the source localization with networks of GW detectors is in the focus of research
in the gravitational wave data analysis. There are several analytical 
studies~\cite{triang, triang2008, Feir, WenChen} of this
problem considering geometrical reconstruction of source coordinates based on the triangulation,
which requires a  measurement of the arrival time of a GW signal at different detectors. 
However, the accurate timing of the GW signal  is intimately related to the reconstruction
of the signal waveforms.  
Due to the different detector sensitivities to the GW polarizations,
the waveforms recorded by individual detectors may be different and they may not
have a common timing reference (like a signal peak time) for a direct measurement of the 
differences in the arrival time. 
Therefore, the problem of the source localization is better addressed in the framework of 
the coherent network analysis~\cite{GT,FH,PRD05}, which reconstructs the waveforms
and the sky coordinates simultaneously.
By using both these methods (triangulation and coherent network analysis), several practical 
source localization algorithms~\cite{klimenko2008, Omega, MBTA} have been recently developed 
and used during the LIGO and Virgo data taking runs in 2009-2010.

There have been a number of studies addressing benefits of individual
detectors~\cite{Schutz1987,Searl2006} and various detector networks~\cite{Weiss2010,Schutz2010}.
In this paper we present a simulation study of the source localization and 
the reconstruction of GW waveforms with the networks of advanced detectors. 
The study is performed with a coherent network method, called coherent 
WaveBurst~\cite{klimenko2008} (cWB), based on the likelihood analysis 
.
In cWB the data from all detectors in the network is processed simultaneously in order 
to reconstruct a common GW signal which is consistent with the recorded detector responses.
The consistency is measured by the likelihood ratio, which is a function of the source
parameters (waveforms and sky location).  The most probable source parameters are obtained 
by maximizing the likelihood ratio over the signal waveforms and sky coordinates.
The method performs reconstruction of unmodeled burst 
signals (arbitrary waveforms) and signals with a certain polarization state:
elliptical, linear and circular. 

The paper is organized as follows. 
Possible networks of advanced detectors and their fundamental properties are descussed 
in section~\ref{sec:networks}. 
In sections~\ref{sec:algorithm} we describe the reconstruction 
algorithm. The simulation framework for this study is presented in section~\ref{framework}. 
The results are reported in section~\ref{Results}. In sections~\ref{limitation} 
and~\ref{conclusions} we 
describe main factors limiting the source reconstruction and discuss the results.

\section{Detector Networks}\label{sec:networks}
\label{network}

In 2001-2010 the LIGO Scientific Collaboration (LSC) and the Virgo Collaboration
operated a network of interferometric gravitational-wave detectors 
which are the most sensitive instruments from the 
first generation of the GW interferometers (1G).
They consist of power-recycled Michelson interferometers with
kilometer-scale Fabry-Perot arms designed to detect gravitational waves
with frequencies between tens of Hz and several kHz. 
The two LIGO observatories~\cite{LIGO}
are in Hanford, Washington (4~km and 2~km detectors) 
and in Livingston, Louisiana (4~km detector), 
and the 3-km Virgo detector~\cite{Virgo} is located in Cascina, Italy.
Other gravitational waves interferometers are the 300 m detector 
TAMA~\cite{TAMA} in Mitaka, Japan, 
and the 600 m detector GEO600~\cite{GEO} in Hannover, Germany. 
Currently all 1G interferometers are decommissioned, except Virgo and GEO600,
which continue to take data. 

\begin{figure}[!hbt]
 \begin{center} 
  \begin{tabular}{c}
 \includegraphics[width=80mm]{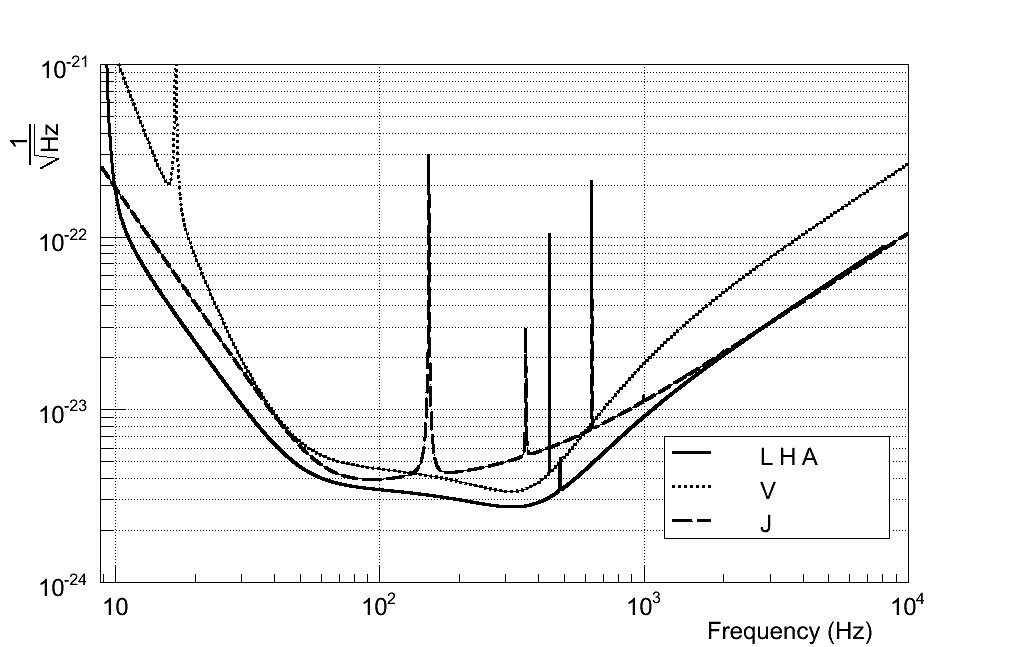}
    \end{tabular}
 \end{center}\caption{\small{\textit{Amplitude spectral density of the design noise for 
the second generation detectors.}}}
\label{Fig:2Gsensitivities}
\end{figure}

The second-generation GW detectors (2G) are currently under construction. 
They include the advanced
LIGO detectors~\cite{advLIGO}, and the advanced 
Virgo detector (V)~\cite{advVirgo} which will have
by an order of magnitude better sensitivity than the 1G detectors.   
All advanced LIGO detectors have 4-km long arms,  with one detector 
in Livingston (L) and two identical co-aligned detectors in Hanford (H and \~{H}).
Also there are plans to build the Large Cryogenic Gravitational Telescope (LCGT)
~\cite{LCGT,LCGT2010} in Japan (the J detector) and possibly move the LIGO \~{H} detector 
to a site in Australia~\cite{AIGO,Weiss2010} (the A detector).
Figure~\ref{Fig:2Gsensitivities} shows the design sensitivity for the
listed 2G detectors.
These, hopefully all five interferometers, compose the most advanced GW detector 
network which will be in operation after 2015.  

The network performance greatly depends on the number of detectors in the network, 
their location and orientation of the detector arms.  
Table~\ref{location} shows the geographical coordinates of the instruments and
the orientation of the detector arms used in this study.
\begin{table}
\centering
\begin{tabular}{|c|c|c|c|}\hline
detector & latitude & longitude & orientation\\
\hline
L & 30$^{\circ}$33$'$46$''$N & 90$^{\circ}$46$'$27$''$W & 197.7 \\
\hline
H & 46$^{\circ}$27$'$18$''$N & 119$^{\circ}$24$'$27$''$W & 126.0 \\
\hline
V & 43$^{\circ}$37$'$53$''$N & 10$^{\circ}$30$'$16$''$E & 70.6 \\
\hline
A & 31$^{\circ}$21$'$30$''$S & 115$^{\circ}$42$'$30$''$E & 45.0 \\
\hline
J & 36$^{\circ}$15$'$00$''$N & 137$^{\circ}$10$'$48$''$E & 19.0 \\
\hline
\end{tabular}
\caption{\small{\textit{Geographical locations and orientations of the 2G detectors.
The orientation of the detector arms is defined  by the rotation angle (counterclockwise)
with respect to the local coordinate frame with axises due North and East.}}}
\label{location}
\end{table}
For the Australian instrument the orientation of the detector arms is not yet decided,
therefore we consider two possible configurations:
\~{A} - arms are due north and east, and 
A - arms are rotated counterclockwise by $45^o$ with respect to \~{A}.
Depending what instruments are constructed we consider several network 
configurations, as listed in Table~\ref{table:SV}.

\subsection{Network sensitivity}

The sensitivity of the network of $K$ detectors is fully characterized by 
its noise-scaled antenna pattern vectors  ${\bf{f_+}}$ and ${\bf{f_{\times}}}$:
\begin{equation}
\label{eq:apn}
{\bf{f_{+(\times)}}} = \left( \frac{F_{1+(\times)}} 
{{\sigma_1}},..,\frac{F_{k+(\times)}}{{\sigma_k}},..,\frac{F_{K+(\times)}}{{\sigma_K}} \right) \;,
\end{equation}
where $\sigma^2_k$ is the variance of the noise and ($F_{k+},F_{k\times}$) are 
the antenna patterns of individual detectors. In general, 
the $\sigma_k$ is a function of frequency. For a discrete data with 
the Nyquist frequency $f_N$ in the Fourier or wavelet domain, the $S_k=\sigma^2_k/f_N$ 
is the estimator of the single-sided power spectral density of the detector noise.
The power spectral density $S_{\mathrm{net}}$ and the variance $\sigma^2_{\mathrm{net}}$
of the network noise are defined by the following equation:
\begin{equation}
\label{eq:snet}
 S_{\mathrm{net}} = \frac{\sigma^2_{\mathrm{net}}}{f_N} = 
\left(f_N \sum_{k=1}^K{\sigma^{-2}_k}\right)^{-1} \;.
\end{equation}
Note, that both the $S_{\mathrm{net}}$ and $\sigma^2_{\mathrm{net}}$ decrease
as more detectors are added to the network. A network of K equally sensitive
detectors has by a factor $\sqrt{K}$ lower noise amplitude 
than the individual detectors. 

The antenna patterns depend upon the source coordinates
($\theta,\phi$) and the polarization angle $\Psi$, which defines the wave
frame of the incoming GW signal ${\bf{h}}=(h_+,h_\times)$. 
It is convenient to define vectors ${\bf{f_+}}$, ${\bf{f_{\times}}}$ and ${\bf{h}}$ 
in the dominant polarization wave frame~\cite{PRD05} 
where $({\bf{f_+}} \cdot \bf{f_{\times}})=0$ and $|{\bf{f_{+}}}| \ge |{\bf{f_{\times}}}|$. 
The vectors ${\bf{f_+}}$ and ${\bf{f_{\times}}}$
define a vector of the noise-scaled detector 
responses to the wave ${\bf{h}}$
\begin{equation}
\label{eq:xih}
{\bf{\xi_{\mathrm{h}}}} = {\bf{f_{+}}}h_+ + {\bf{f_{\times}}}h_{\times} \;.
\end{equation}
The inner product $({\bf{\xi_{\text{h}}}}|{\bf{\xi_{\text{h}}}})$
calculated over the sampled detector responses 
gives the estimator of the network signal-to-noise ratio (SNR):
\begin{equation}
\label{eq:RHO}
\rho_{\mathrm{net}}=\sqrt{({\bf{\xi_{\text{h}}}}|{\bf{\xi_{\text{h}}}})}.
\end{equation}
In general, the inner product of two network vectors {\bf{a}} and {\bf{b}} is defined as
\begin{equation}
\label{eq:inner}
({\bf{a}}|{\bf{b}}) = \sum_i{({\bf{a}}[i] \cdot {\bf{b}}[i])} 
\end{equation}
where the sum is taken over the data samples $i$ containing the event.

The norms of the antenna pattern vectors 
$|{\bf{f_{+}}}|$ and $|{\bf{f_{\times}}}|$ characterize the network sensitivity to 
the GW polarizations. 
To illustrate this and other network properties, we assume below that in 
the signal frequency band  the vectors  ${\bf{f_+}}$ and ${\bf{f_{\times}}}$ do
not vary much.  In this case 
\begin{equation}
\label{eq:rho}
\rho_{\mathrm{net}} \approx \sqrt{|{\bf{f_+}}|^2 (h_+|h_+) + |{\bf{f_{\times}}}|^2 (h_{\times}|h_{\times})}
\end{equation}
where the inner product $({\bf{h}}|{\bf{h}})=(h_+|h_+)+(h_{\times}|h_{\times})$ 
determines the root-sum-square amplitude of the GW polarizations:
\begin{equation}
\label{eq:hrss}
h_{\text{rss}} = \sqrt{\frac{({\bf{h}}|{\bf{h}})}{2f_N}} \;. 
\end{equation}

As it follows from Equation~\ref{eq:rho}, the network alignment factor~\cite{klimenko2008} 
\begin{equation}
\label{eq:alp}
\alpha = |{\bf{f_{\times}}}|/|{\bf{f_{+}}}| 
\end{equation}
characterizes the relative network sensitivity to the two GW polarizations. 
It determines the ratio of the  SNRs from each GW component, assuming that in average
their sum-square energies are the same: $(h_+|h_+)=(h_{\times}|h_{\times})$.
Closely aligned networks (like LH\~{H}) have poor sensitivity to 
the second polarization ($\alpha \ll 1$) making difficult the 
reconstruction of the full GW signal.

\begin{figure*}[!hbt]
\begin{center} 
\begin{tabular}{cc}
\includegraphics[width=0.45\textwidth]{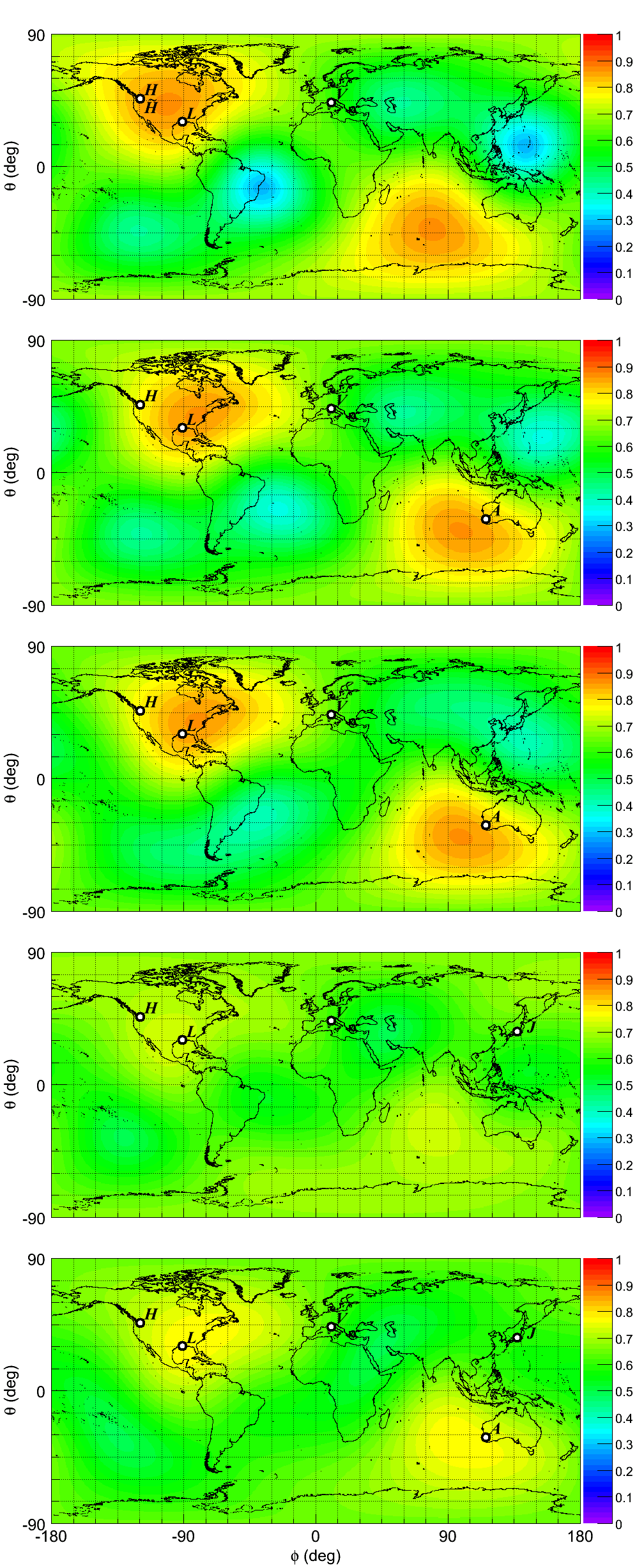} &
\includegraphics[width=0.45\textwidth]{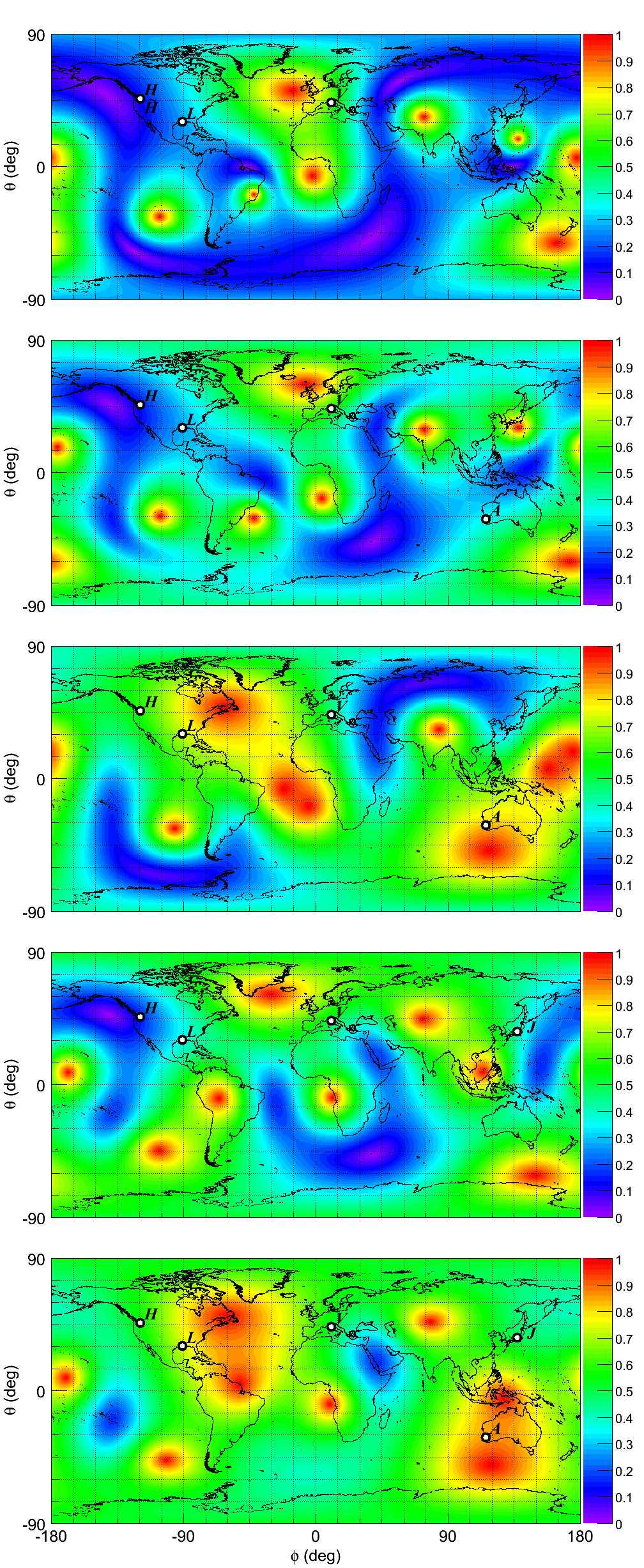}
\end{tabular}
\end{center}
\caption{\small{\textit{The distributions of ${\cal F}$ 
(left plots) and $\alpha$ (right plots) at the frequency of 100~Hz as a function of latitude ($\theta$)
and longitude ($\phi$) for the networks LH\~{H}V, LHV\~{A}, LHVA, LHVJ, LHVAJ (from top to bottom). }}}
\label{Fig:fpfx}
\end{figure*}

The overall network sensitivity is characterized by 
the effective power spectral density of the network noise 
\begin{equation}
\label{eq:Nnet}
N_{\text{net}} = f^{-1}_N \left(|{\bf{f_+}}|^2 + |{\bf{f_{\times}}}|^2\right)^{-1} 
\end{equation}
which depends on the sky coordinates. It determines the average network SNR for a population
of GW signals with the average amplitude $h_{\text{rss}}/\sqrt{2}$ per polarization: 
\begin{equation}
\label{eq:avrho}
\overline{\rho}_{\mathrm{net}} 
\approx \frac{h_{\text{rss}}}{\sqrt{N_{\text{net}}}} \;.
\end{equation}
It is convenient to factorize the sky-dependent part of $N_{\text{net}}$ as
\begin{equation}
\label{eq:factorize}
N_{\mathrm{net}} = {\cal F}^{-2} S_{\text{net}} \;,
\end{equation}
where ${\cal F}$ is the network antenna factor distributed
between 0 (low sensitivity) and 1 (high sensitivity).
Figure~\ref{Fig:fpfx} shows the antenna and the alignment factors 
for different networks as a function of the latitude and longitude of 
the source (skymaps). Since these network parameters are noise dependent, 
the skymaps are calculated at the frequency 100~Hz where the advanced detector
sensitivities are about the same.
The ${\cal F}$ distribution shows how uniform is the network response across the sky.
The values of $\alpha$ close to unity indicate the same sensitivity 
to the two GW components. Respectively, the values of $\alpha$ close to zero indicate 
that the second GW component is not measurable for a weak GW signal. 
As Figure~\ref{Fig:fpfx} shows, several non-aligned 
detectors (preferably five) are required for elimination of these blind spots in the sky.

One of the main characteristics of a detector network is its 
search volume. Given an isotropic distribution of transient sources with
the root-square-sum amplitude $h_o$ at the fiducial distance $r_o$, the search volume
is defined as~\cite{S5-VSR1-y2}
\begin{eqnarray}
 V_{\text{net}} = 4\pi (h_o r_o)^3 \int_0^\infty 
 \!\! dh \, h^{-4} \epsilon(h) \, .
\end{eqnarray}
where $\epsilon$ is the detection efficiency. Assuming the same SNR thresholds 
$\rho_{\text{net}}(h)$ (see Equation~\ref{eq:avrho}),
the  $V_{\text{net}}$ can be calculated with respect to the volume 
$V_0$ of the reference network
\begin{eqnarray}
{V_{\text{net}}} = V_0 \frac{<N^{3/2}_{0}>}{<N^{3/2}_{\text{net}}>} 
\end{eqnarray}
where $<N^{3/2}_{\text{net}}>$ and $<N^{3/2}_{0}>$ are 
the averages over the sky. 
The ratio $V_{\text{net}}/V_0$ is quite independent on the search algorithm and the GW source model.
Table~\ref{table:SV} shows the volume (and detection rate) ratios calculated with respect 
to the LHV network ($V_0=V_{\text{LHV}}$).
\begin{table}
  \center
  \begin{tabular}{|c|c|c|c|c|c|}\hline
        & LHV & LH\~{H}V & LHVA & LHVJ & LHVAJ \\ \hline
100~Hz & 1 & 1.66 & 1.65 & 1.39 & 2.09 \\ \hline
300~Hz & 1 & 1.65 & 1.63 & 1.15 & 1.80 \\ \hline
\end{tabular}
\caption{\small{\textit{Expected difference in detection rates with respect to the LHV network.}}}
\label{table:SV}
\end{table}
As more detectors are added to the network the detection rates increase.
Being beneficial this increase, however, is not critical
for the first direct observation of gravitational waves.
More important, the networks with more detectors are likely to be less affected 
by non-Gaussian and non-stationary noise than the networks with fewer detectors. 
They are expected to have lower false alarm rates and higher detection confidence
for the same $\rho_{\text{net}}$ threshold. 
Also, re-location of the LIGO \~{H} detector to Australia does not affect the detection
rates for both A and \~{A} configurations. But, as shown in Figure~\ref{Fig:fpfx}, 
the LHVA detector configuration would be more preferable for the reconstruction of 
the GW polarizations than the LH\~{H}V and LH\~{A}V networks.

\section{Reconstruction Algorithm}
\label{sec:algorithm}
\subsection{Coherent network analysis}

One possible approach to the coherent network analysis is based on the Neyman-Pearson criterion
which defines the likelihood ratio
\begin{equation}
\label{eq:LIKE}
\Lambda({\rm x,\Omega}) = \frac{p({\rm x}|{\bf{h}}(\Omega))}{p({\rm x}|0)}\;,
\end{equation}
where $x$ is the network data vector, the $p({\rm x}|0)$ is the joint probability that 
the data is only instrumental noise and $p({\rm x}|{\bf{h}})$ is the joint probability that
a GW signal ${\bf{h}}$ is present in the data. In general the likelihood ratio is 
a functional which depends upon the source parameters $\Omega$. 
One generalization of the Neyman-Pearson criterion 
is to maximize $\Lambda({\rm x},\Omega)$ over $\Omega$. The obtained 
maximum likelihood ratio statistic reaches its maximum for the best match of the corresponding 
waveform to the data. If the source model allows calculation of the GW waveforms as 
a function of a small number of source parameters 
(for example, for binary black holes), then a template bank can be generated. 
In this case the variation is performed over the template bank and the likelihood 
approach is equivalent to a matched filter.
The cWB algorithm searches for unmodeled burst signals.
In contrast to the binary black hole sources, where the number of parameters is 
relatively small, 
the parameters characterizing the unmodeled bursts are essentially the signal amplitudes 
themselves at each instance of time. It is not possible to generate a template bank for
such a large parameter space. Instead, the best matching waveform is found
by variation of $\Lambda$ over unknown GW waveforms ${\bf{h}}$.

\subsection{GW waveforms}
For stationary Gaussian noise 
the coherent WaveBurst algorithm defines the likelihood ${\cal L}$ 
as twice the logarithm of the likelihood ratio $\Lambda$
\begin{equation}
\label{eq:like}
{\cal{L}}[{\bf h}] = 2({\bf{w}} \vert {\bf{\xi_{\text{h}}}})- ({\bf{\xi_{\text{h}}}} \vert {\bf{\xi_{\text{h}}}}) \, ,
\end{equation}
where the vector ${\bf{w}}$ represents whitened data from $K$ detectors 
with uncorrelated noise
\begin{equation}
\label{eq:vecW}
{\bf{w}} = \left( \frac{x_1[i,\tau_1]}{\sigma_1},..,\frac{x_k[i,\tau_k]}{\sigma_k}
,..,\frac{x_K[i,\tau_K]}{\sigma_K} \right) \;.
\end{equation}
The sampled detector amplitudes 
$x_k[i,\tau_k]$ take into account the time-of-flight delays $\tau_k$, 
which in turn depend upon the source coordinates $\theta$ and $\phi$.
The solutions for the GW waveforms ${\bf{h}}$, defined in the dominant polarization frame, 
are found by the variation of the likelihood functional (Eq.~(\ref{eq:like})):
\begin{eqnarray}
\label{eq:syst1}
 {\it H}_+ = {({\bf{w}}\cdot{\bf{f_+}})}/|{\bf{f_+}}|^2 \;, \\
\label{eq:syst2}
 {\it H}_\times = {({\bf{w}}\cdot{\bf{f_\times}})}/|{\bf{f_\times}}|^2 \;. 
\end{eqnarray}
The maximum likelihood ratio statistic is calculated by 
substituting the solutions into ${\cal L}[{\bf h}]$. The result can be written as
\begin{equation}
\label{eq:lMax}
L_{\mathrm{max}} = \sum_i {\bf{w}} P {\bf{w}}^T \,,
\end{equation}
where the matrix $P$ is the projection constructed from the components of the 
unit vectors ${\bf{e_+}}$ and ${\bf{e_\times}}$ along the directions of 
the ${\bf{f_+}}$ and ${\bf{f_\times}}$ respectively:
\begin{equation}
P_{nm}=e_{+n}e_{+m}+e_{\times{n}}e_{\times{m}} \, .
\end{equation}
The kernel of the projection $P$ is the {\it signal} plane defined by these two vectors. 
The null space of the projection $P$ 
defines the reconstructed detector noise $n_{\text{rec}}$ which is referred to as the null stream.

The projection matrix is invariant with respect to the rotation in the signal plane where
any two orthogonal unit vectors can be used for construction of the $P_{nm}$.
Therefore one can select unit vectors ${\bf{u}}$ and ${\bf{v}}$ such that
${\bf{w}}\cdot{\bf{v}}=0$ and then
\begin{equation}
\label{eq:P(u)}
P_{nm}=u^nu^m \, .
\end{equation}
The unit vector ${\bf{u}}$ defines the vector 
\begin{equation}
\label{eq:xi}
{\bf{\xi}} = ({\bf{w}} \cdot {\bf{u}}){\bf{u}}
\end{equation}
whose components are the standard likelihood estimators  of 
the noise-scaled detector responses $\xi^k_{\text{h}}$.

\subsection{Source coordinates}
The maximum likelihood ratio statistic $L_{\mathrm{max}}$ is a function 
of the sky coordinates $\theta$ and $\phi$. If no information regarding the source 
coordinates is available then the variation over the sky should be also performed. 
It is expected that $L_{\mathrm{max}}$ takes maximum close to a true source
location, however, it is not necessarily the optimal statistic for the coordinate
reconstruction. The coherent part of the likelihood quadratic form
\begin{equation}
E_c =  \sum_i {\sum_{n,m}{w_nw_mP_{nm}},~~~n\neq{m}}
\end{equation}
has a strong dependence on the time delays between the detectors and therefore the 
coherent energy $E_c$ is expected to be a better statistics for the source localization.
On the other hand the $E_c$ is a biased estimator: for an arbitrary 
GW signal it may take maximum value away from the true source location.
To minimize the bias, the sky statistic is constructed in the following way
\begin{equation}
L_{\mathrm{sky}}  = \frac{L_{\mathrm{max}}E_c}{E(E-L_{\mathrm{max}}+|E_c|)}
\end{equation}
where $E = (\bf{w}|{\bf{w}})$ is the total normalized energy of the signal and noise.
This statistic penalizes the sky locations with low values of $E_c$ and large values of
the null energy $E-L_{\mathrm{max}}$, and it reduces to the maximum likelihood statistic 
$L_{\mathrm{max}}/E$ when the ratio $(E-L_{\mathrm{max}})/E_c$ is small. 
The $L_{\mathrm{sky}}$ is used to rank different sky locations and calculate the probability 
distribution of the estimated source coordinates in the sky.

\subsection{Model-dependent constraints}

The likelihood method offers a convenient framework for introduction of
constraints arising from the source models.  Unlike for template searches where 
accurate waveforms are required, in principle, any useful information about sources can 
be used to constraint the likelihood functional.
This allows customization of the generic burst algorithms in order  to search for specific, 
but not very well modeled sources. One obvious class of constraints is related to the
different polarization states of the GW signals. For example, some of the core collapse
models predict waveforms with a linear polarization~\cite{SN}
or random polarization~\cite{g-modes Ott}.  
Merging binary neutron stars or black holes 
are expected to produce elliptically polarized gravitational wave signals~\cite{Pankow}.
Also the neutron star mergers can be the source of the short GRB 
signals~\cite{grb070201} where relativistic jets
are emitted along the rotation axis of the binary system and in this case
the associated gravitational waves should have the circular polarization.
The cWB algorithm allows searches with several types of the polarization
constraints: circular, linear, elliptical and random (or unmodeled search). 
All these searches are used in the study to estimate possible improvement of
the source localization if the reconstruction is constrained by the source model.

\section{Simulations}\label{framework}
\subsection{Injected signals}\label{sec:Simulated signals}
Several types of ``ad hoc'' waveforms were used to study the performance of 
the detector networks for different signal frequencies and polarization states.
They were injected into the simulated detector data streams in a wide range of signal-to-noise
ratios with the coordinates uniformly distributed in the sky. 
The Gaussian detector noise was simulated with the amplitude spectral density
presented in Figure~\ref{Fig:2Gsensitivities}.
The injected signals were band-limited white-noise waveforms (WNB)
with the random polarizations and sine-Gaussian waveforms (SG) with 
the linear and circular polarizations.
The WNB waveforms were bursts of white Gaussian noise in a frequency band $(f_1,f_2)$ 
which have a Gaussian time profile with the standard deviation $\tau$ (see Table~\ref{Tab: injWNB}).
The random polarization waveforms $h_+$ and $h_{\times}$ were selected to have the same
square-sum energy: $(h_+|h_+) = (h_{\times}|h_{\times})$

\begin{table}[|h!btc]
\begin{center}
\begin{tabular}{|c|c|c|c|}
\hline
 Waveform & $\tau$ (s) & $f_1$ (Hz) & $f_2$ (Hz)\\
 \hline
 WNB & 0.1 &   250 &   350 \\
 \hline
 WNB & 0.1 &   1000 &   2000 \\
 \hline
 \end{tabular}
\end{center}
\caption{\small{\textit{Simulated White Noise Bursts.}}}\label{Tab: injWNB}
\end{table}

The SG waveforms were simulated as follows
\begin{eqnarray}
h_+(t)= h_o \sin(2 \pi t f_0)\exp(-t^2/\tau^2) \;, \\
h_{\times}(t)= h_1 \cos(2 \pi t f_0)\exp(-t^2/\tau^2) \;,
\end{eqnarray}
where $f_0$ is the waveform central frequency, $h_o$ and $h_1$ are the waveform
amplitudes, and $\tau$ is related to the 
waveform quality factor $Q=\sqrt{2} \pi f_0 \tau$ (see Table~\ref{Tab: injSG}). 
The amplitude parameters  were $h_1=0$ for linear polarization and 
$h_1=h_o$ for circular polarization. During the analysis the amplitude of
injected events varied to simulate events with different signal-to-noise ratios.
\begin{table}[|h!btc]
\begin{center}
\begin{tabular}{|c|c|c|c|}
\hline
 Waveform & $f_0$ (Hz) & Q  & Polarization\\
 \hline
 SGQ3 &   235/1053  &   3 & Linear \\
 \hline
 SGQ9         &   235/1053  &   9 & Linear\\
 \hline
 SGCQ9      &   235/1053 &    9  & Circular\\
 \hline
 \end{tabular}
\end{center}
\caption{\small{\textit{Simulated sine-Gaussian waveforms with quality factors Q=3 and Q=9, 
low (235Hz) and high (1053Hz) frequencies, and two polarization types - linear and circular.}}}
\label{Tab: injSG}
\end{table}
 
\subsection{Error regions}\label{skystat}

\begin{figure}[!hbt]
 \begin{center} 
  \begin{tabular}{c}
 \includegraphics[width=80mm]{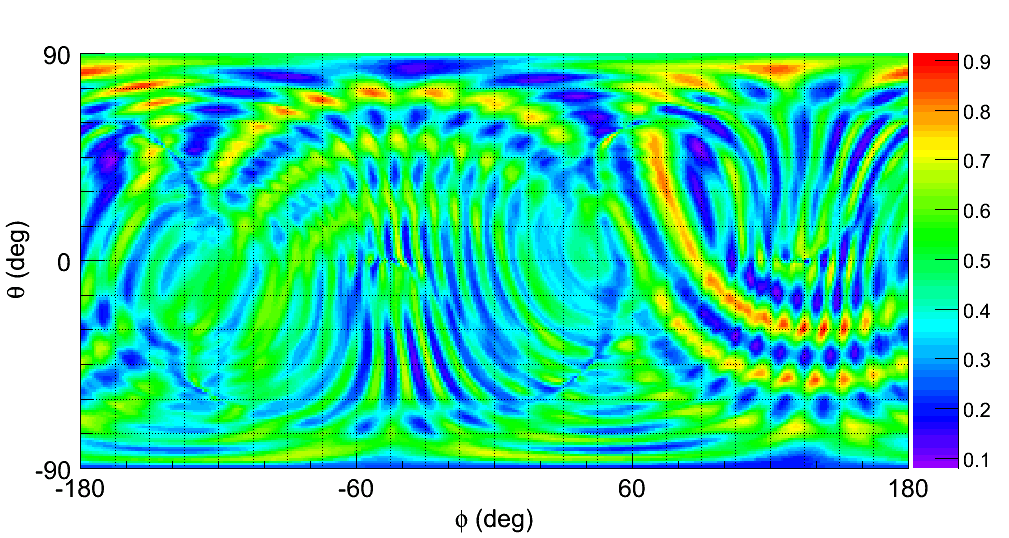}\\
 \includegraphics[width=80mm]{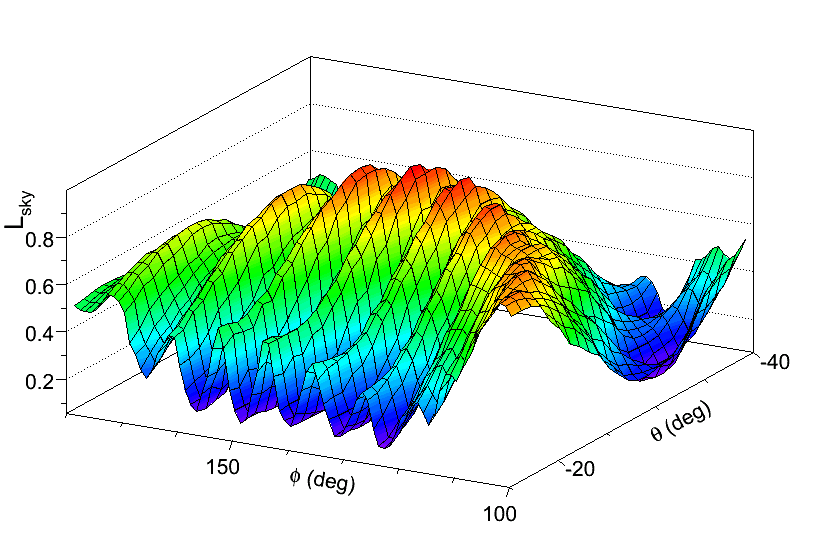}
   \end{tabular}
 \end{center}\caption{\small{\textit{Example of the likelihood sky map $L_{\mathrm{sky}}$  
for an injected signal at $\theta=-30^\circ$ and $\phi=144^\circ$: $L_{\mathrm{sky}}$ as a function 
of $\theta$ and $\phi$ (top), $L_{\mathrm{sky}}$ distribution around the reconstructed 
location (bottom).}}} 
\label{Fig. Like1evt}
\end{figure}

The injected signals are used for estimation of the accuracy of the coordinate reconstruction. 
For each injected event the $L_{\mathrm{sky}}$ skymaps is calculated with the angular 
resolution of $d\Omega=0.4\times0.4$ square degrees: $\sim200,000$ sky locations (pixels) total.
Figure~\ref{Fig. Like1evt} shows an example of such a skymap for one of the SGQ9 (235Hz) injections. 
\noindent
For such events it is typical to see a pattern of fringes with large value of 
$L_{\mathrm{sky}}$ corresponding to a good match between responses due to 
a common GW signal reconstructed in different detectors. Such sky points are the most probable 
as the source location. Depending on many factors, such as the signal strength, waveform morphology, etc,
the $L_{\mathrm{sky}}$ statistic can be well localized in a single small cluster in the sky
or distributed over a large area which can be also split into several disjoint clusters.
This type of ambiguity is typical for the least constrained unmodeled search
and networks with only  three spatially separated detectors.

To characterize the accuracy of the coordinate reconstruction for a single injection
we define an error region: 
total area of all pixels in the sky which satisfy the condition 
$L_{\mathrm{sky}}(\theta,\phi)\geq{L_{\mathrm{sky}}(\theta_i,\phi_{i})}$,
where ($\theta_i,\phi_{i}$) is the injection sky location. 
Given a population of injected signals uniform in the sky, the 50~CL and 90~CL error regions,
containing $50\%$  and $90\%$ of injections respectively, can be calculated.
The median error angle is defined as the square root of the $50\%$ error area.

The $L_{\mathrm{sky}}$ skymap can be also converted into the probability skymap 
which is normalized to unity if integrated over the entire sky. In this case  
the 50~CL and 90~CL error regions are represented by the most probable pixels with the
cumulative probability of $50\%$  and $90\%$ respectively. 
Such probability skymaps are not relevant for the simulation studies we
perform, but they are important for the analysis of real data.

\section{Results}\label{Results} 

\subsection{Coordinate Reconstruction}\label{Res:CoorRec}

\begin{figure}[!hbt]
 \begin{center} 
 \includegraphics[width=0.5\textwidth]{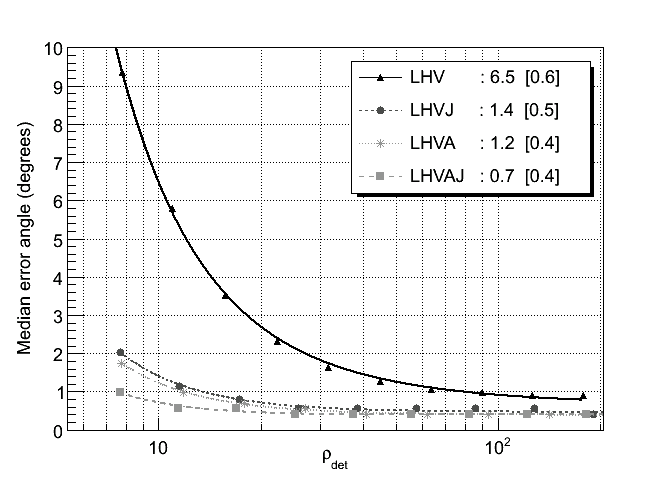}
 \end{center}\caption{\small{\textit{Median error angle vs average detector SNR obtained with 
the unmodeled algorithm for all types of injections and different network configurations: 
LHV, LHVA, LHVJ, LHVAJ.}}}
 \label{MAvsSNR}
\end{figure}

The accuracy of the coordinate reconstruction strongly depends on the strength of detected signals
which can be conveniently characterized by the average (per detector) signal-to-noise ratio
\begin{equation}
\rho_{\text{det}} = \rho_{\text{net}}/\sqrt{K}  \;.
\end{equation}
For example, Figure~\ref{MAvsSNR} shows the dependence of the median error angle $\alpha_{50\%}$ 
on $\rho_{\text{det}}$ for all injected signals, which is well approximated
by a function
\begin{equation}
\alpha_{50\%}=A+B \left(\frac{10}{\rho_{\text{det}}} \right) + 
C \left(\frac{10}{\rho_{\text{det}}} \right)^2\;.  
\end{equation}
The parameter A is the median error angle for events with very large SNR and 
A+B+C is the median error angle for events with $\rho_{\text{det}}=10$. 
\noindent
Figure~\ref{MAvsSNR}  also shows a dependence of the coordinate resolution on 
the number of detector sites in the network. 
There is a significant improvement of the resolution when more sites are 
added to the network. This is particularly noticeable at low SNR,
which is very important because the anticipated GW signals are likely to be weak.

\begin{figure}[!hbt]
\begin{center} 
\includegraphics[width=80mm]{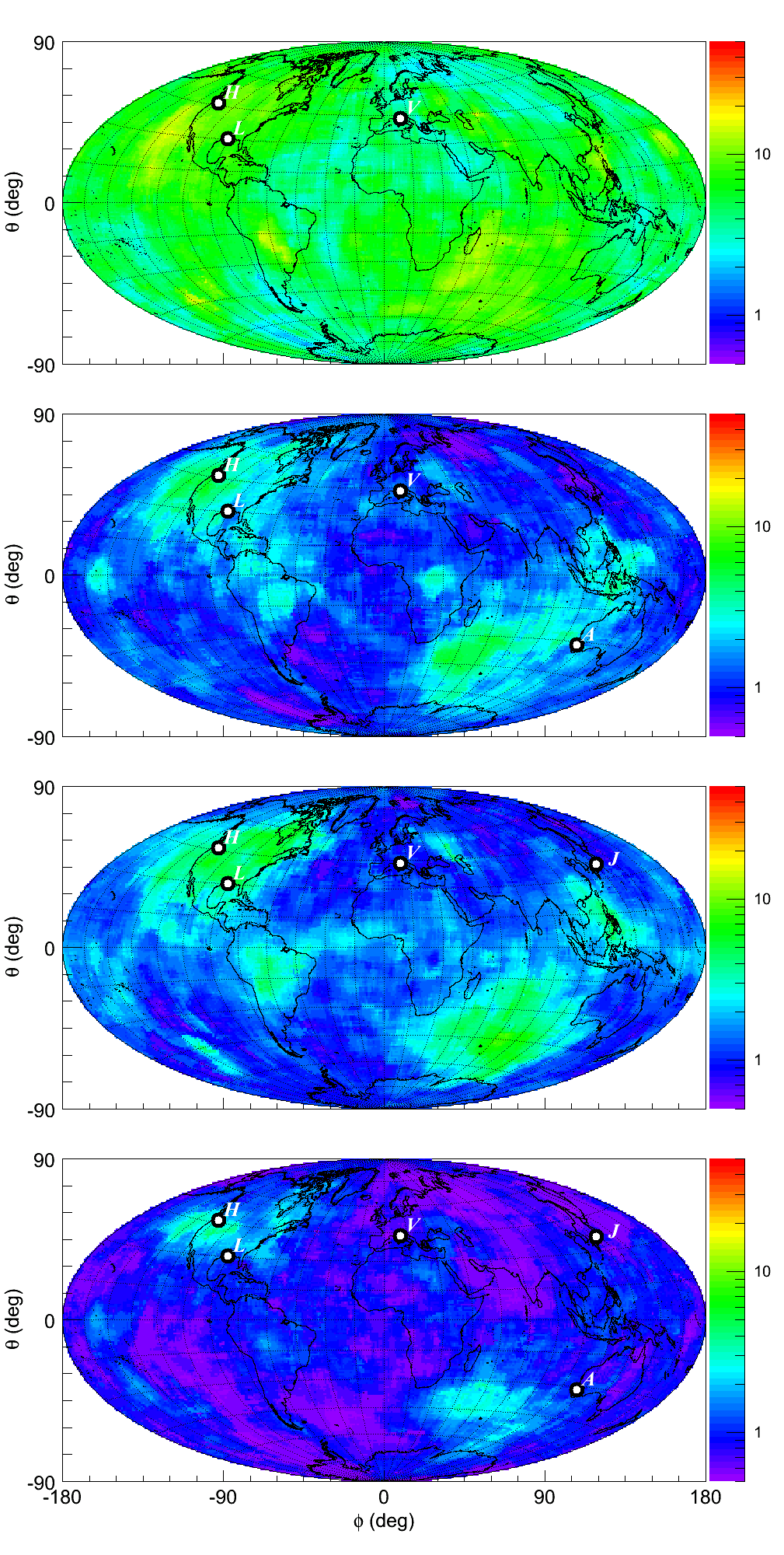}
\end{center}\caption{\small{\textit{Median error angle for LHV, LHVA, LHVJ and LHVAJ networks
(from top to bottom) as a function of source coordinates ($\theta$ - latitude, $\phi$ - longitude)
for injections with the network SNR between 10 and 30.}}}
\label{sky50maps}
\end{figure}

\begin{figure}[!hbt]
\begin{center} 
\includegraphics[width=80mm]{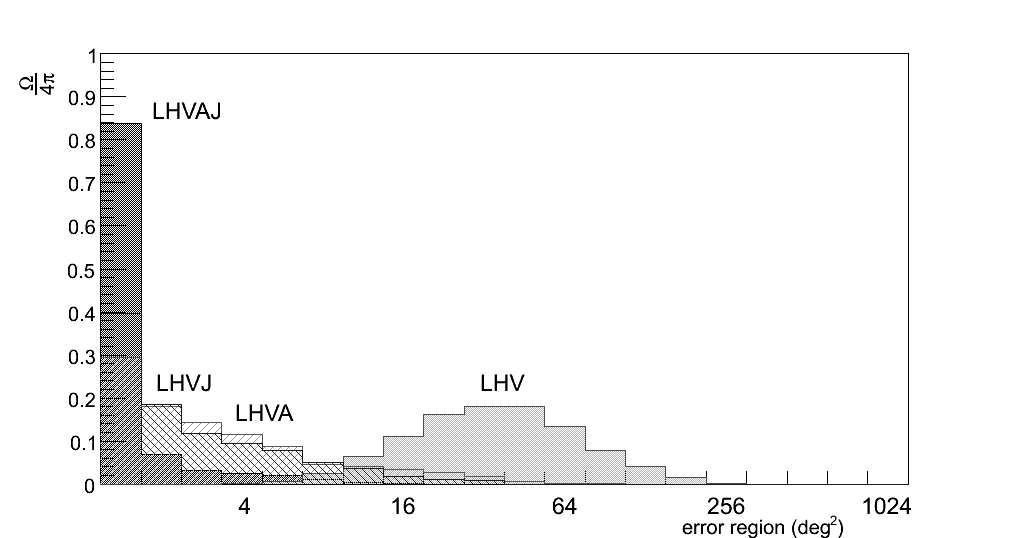}
\includegraphics[width=80mm]{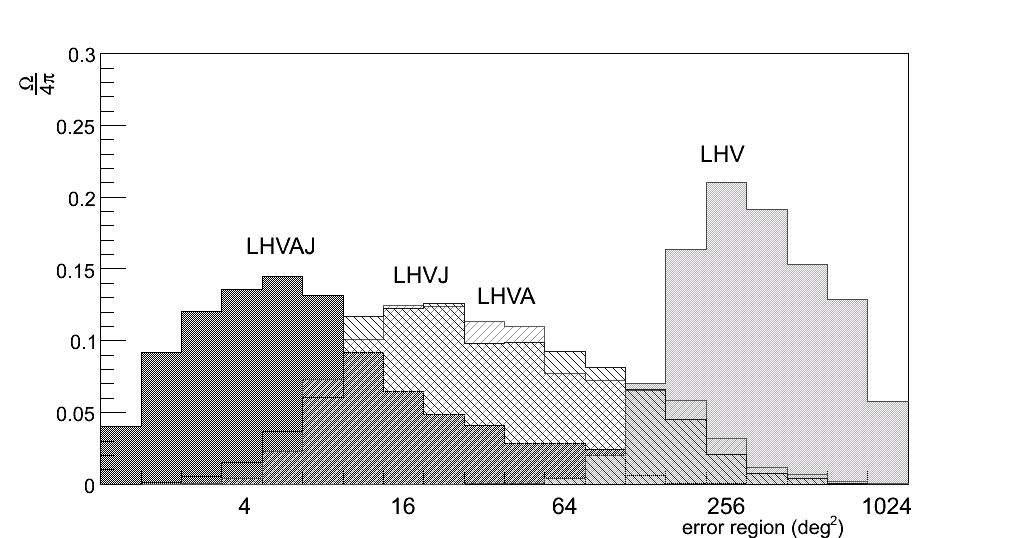}
\label{Lars}
\end{center}\caption{\small{\textit{Fraction of the sky (vertical axis) for 3-site (LH\~{H}V), 
4-site (LHVA and LHVJ) and 5-site (LHVAJ) networks where sources are reconstructed
with a given  $50\%$ (top)
and $90\%$ (bottom) error region (horizontal axis).}}}
\label{LarsHistograms}
\end{figure}

Because of several limiting factors
(see section~\ref{limitation}) the reconstruction is not uniform in the sky.
Figure~\ref{sky50maps} shows the distribution of the
median error angle across the sky for different network configurations.
There is a dramatic improvement of the coordinate reconstruction for the 
LHVA, ALVJ and LHVAJ networks. However for the 4-site networks
there remain areas where the source localization is poor.
Figure~\ref{LarsHistograms} compares the pointing capabilities of the network consisting 
of three, four and five sites  by presenting the fraction of the sky where the reconstruction 
is performed with a given error area. This figure also shows a significant improvement of 
the source localization (particularly for the $90\%$ error area) as more sites are used 
for the reconstruction.
The best coordinate resolution is obtained with the 5-site network and 
it is compatible with the field-of-view of most optical telescopes.

\begin{figure}[!hbt]
 \begin{center} 
 \includegraphics[width=0.5\textwidth]{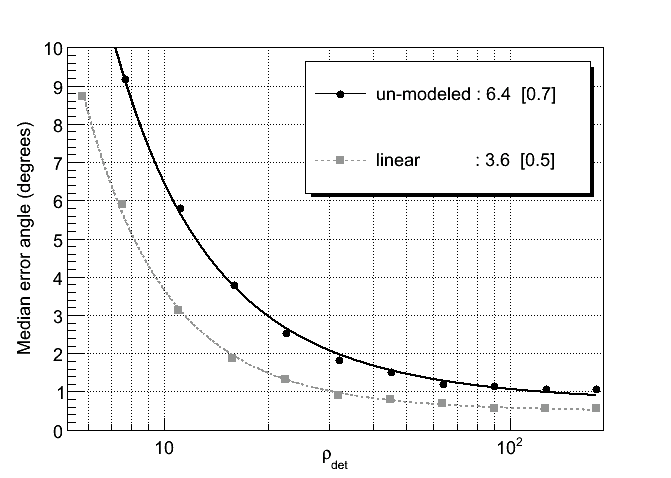}
 \includegraphics[width=0.5\textwidth]{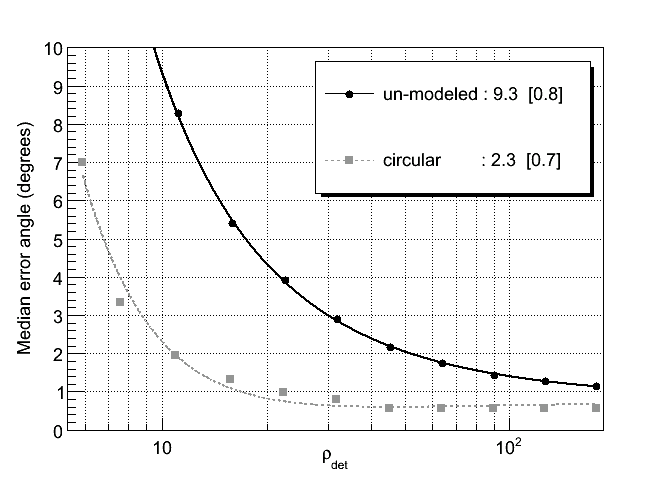}
 \end{center}\caption{\small{\textit{Median error angle vs SNR for LHV network comparing
different constrain searches. Top: SGQ9 LF waveform with unmodeled (black) and linear (red) searches.
Bottom: SGCQ9 LF with unmodeled (black) and circular (red) searches.}}}
 \label{MAvsSNRmod}
\end{figure}

The coordinate resolution depends also on the waveform morphology and the polarization
content of GW signals (see for details Section~\ref{LF_WandP}). If reconstructed 
with the least constrained unmodeled algorithm, the SG waves with linear and circular
polarization have less accurate source localization (see Figure~\ref{MAvsSNRmod}). 
However, the coordinate resolution can be significantly improved if reconstruction is 
constrained by the source polarization model. 
In general, the reconstruction improves as more accurate models, with fewer 
free parameters, are used. We expect,
for example, that the template analysis of waves from the coalescence of binary 
neutron stars and black holes~\cite{cbc2010} should result in a better sky localization 
than for the unmodeled  burst search.

Tables~\ref{Tab:ummodeled}-\ref{Tab:circular} summarize the results of the analysis 
for different injection signals, source polarization models and networks 
by showing the fit parameters A and A+B+C, which correspond  to the median
error angle for events with high and low SNR, respectively.

\begin{table}[|h!btc]
\begin{center}
\begin{tabular}{|c|c|c|c|c|}
\hline
unmodeled & LHV & LHVA & LHVJ & LHVAJ\\
\hline
WNB LF & 4.8$^{\circ}$/0.7$^{\circ}$ & 1.1$^{\circ}$/0.4$^{\circ}$ & 1.8$^{\circ}$/0.4$^{\circ}$ & 0.8$^{\circ}$/0.4$^{\circ}$ \\
\hline
WNB HF & 4.5$^{\circ}$/0.4$^{\circ}$ & 0.6$^{\circ}$/0.4$^{\circ}$ & 0.8$^{\circ}$/0.4$^{\circ}$ & 0.4$^{\circ}$/0.4$^{\circ}$ \\
\hline
SGQ9 LF & 6.4$^{\circ}$/0.7$^{\circ}$ & 1.4$^{\circ}$/0.4$^{\circ}$ & 1.6$^{\circ}$/0.4$^{\circ}$ & 1.0$^{\circ}$/0.4$^{\circ}$ \\
\hline
SGQ9 HF & 4.1$^{\circ}$/0.9$^{\circ}$ & 1.0$^{\circ}$/0.4$^{\circ}$ & 1.0$^{\circ}$/0.4$^{\circ}$ & 0.5$^{\circ}$/0.4$^{\circ}$ \\
\hline
SGQ3 LF & 9.4$^{\circ}$/0.5$^{\circ}$ & 1.1$^{\circ}$/0.5$^{\circ}$ & 1.5$^{\circ}$/0.4$^{\circ}$ & 0.9$^{\circ}$/0.4$^{\circ}$ \\
\hline
SGQ3 HF & 6.3$^{\circ}$/0.4$^{\circ}$ & 0.9$^{\circ}$/0.4$^{\circ}$ & 1.0$^{\circ}$/0.4$^{\circ}$ & 0.5$^{\circ}$/0.4$^{\circ}$ \\
\hline
SGCQ9 LF & 9.3$^{\circ}$/0.8$^{\circ}$ & 1.7$^{\circ}$/0.4$^{\circ}$ & 2.0$^{\circ}$/0.4$^{\circ}$ & 0.9$^{\circ}$/0.4$^{\circ}$ \\
\hline
SGCQ9 HF & 5.5$^{\circ}$/1.1$^{\circ}$ & 1.4$^{\circ}$/0.4$^{\circ}$ & 1.7$^{\circ}$/0.4$^{\circ}$ & 0.9$^{\circ}$/0.4$^{\circ}$ \\
\hline
\end{tabular}
\end{center}
\caption{\small{\textit{Pointing accuracy  (A+B+C / A) for unmodeled search.}}}
\label{Tab:ummodeled}
\end{table}

\begin{table}[|h!btc]
\begin{center}
\begin{tabular}{|c|c|c|c|c|}
\hline
elliptical & LHV & LHVA & LHVJ & LHVAJ\\
\hline
SGQ9 LF & 5.3$^{\circ}$/0.9$^{\circ}$ & 1.2$^{\circ}$/0.4$^{\circ}$ & 1.4$^{\circ}$/0.5$^{\circ}$ & 0.9$^{\circ}$/0.4$^{\circ}$ \\
\hline
SGQ9 HF & 4.5$^{\circ}$/0.8$^{\circ}$ & 0.8$^{\circ}$/0.4$^{\circ}$ & 0.9$^{\circ}$/0.6$^{\circ}$ & 0.5$^{\circ}$/0.4$^{\circ}$ \\
\hline
SGQ3 LF & 3.6$^{\circ}$/0.6$^{\circ}$ & 1.3$^{\circ}$/0.4$^{\circ}$ & 1.1$^{\circ}$/0.4$^{\circ}$ & 0.9$^{\circ}$/0.4$^{\circ}$ \\
\hline
SGQ3 HF & 4.4$^{\circ}$/0.7$^{\circ}$ & 0.9$^{\circ}$/0.4$^{\circ}$ & 0.8$^{\circ}$/0.4$^{\circ}$ & 0.5$^{\circ}$/0.4$^{\circ}$ \\
\hline
SGCQ9 LF & 8.2$^{\circ}$/0.7$^{\circ}$ & 1.9$^{\circ}$/0.4$^{\circ}$ & 1.5$^{\circ}$/0.4$^{\circ}$ & 0.9$^{\circ}$/0.4$^{\circ}$ \\
\hline
SGCQ9 HF & 7.2$^{\circ}$/0.8$^{\circ}$ & 1.4$^{\circ}$/0.4$^{\circ}$ & 1.1$^{\circ}$/0.4$^{\circ}$ & 0.9$^{\circ}$/0.4$^{\circ}$ \\
\hline
\end{tabular}
\end{center}
\caption{\small{\textit{Pointing accuracy  (A+B+C / A) for elliptical search.}}}
\label{Tab:elliptical}
\end{table}

\begin{table}[|h!btc]
\begin{center}
\begin{tabular}{|c|c|c|c|c|}
\hline
linear & LHV & LHVA & LHVJ & LHVAJ\\
\hline
SGQ9 LF & 3.6$^{\circ}$/0.5$^{\circ}$ & 1.1$^{\circ}$/0.4$^{\circ}$ & 1.1$^{\circ}$/0.4$^{\circ}$ & 0.7$^{\circ}$/0.4$^{\circ}$ \\
\hline
SGQ9 HF & 4.5$^{\circ}$/0.6$^{\circ}$ & 0.8$^{\circ}$/0.4$^{\circ}$ & 0.9$^{\circ}$/0.4$^{\circ}$ & 0.5$^{\circ}$/0.4$^{\circ}$ \\
\hline
SGQ3 LF & 2.7$^{\circ}$/0.5$^{\circ}$ & 1.1$^{\circ}$/0.4$^{\circ}$ & 0.9$^{\circ}$/0.4$^{\circ}$ & 0.7$^{\circ}$/0.4$^{\circ}$ \\
\hline
SGQ3 HF & 4.5$^{\circ}$/0.5$^{\circ}$ & 0.8$^{\circ}$/0.4$^{\circ}$ & 0.9$^{\circ}$/0.4$^{\circ}$ & 0.5$^{\circ}$/0.4$^{\circ}$ \\
\hline
\end{tabular}
\end{center}
\caption{\small{\textit{Pointing accuracy  (A+B+C / A) for linear search.}}}
\label{Tab:linear}
\end{table}

\begin{table}[|h!btc]
\begin{center}
\begin{tabular}{|c|c|c|c|c|}
\hline
circular & LHV & LHVA & LHVJ & LHVAJ\\
\hline
SGCQ9 LF & 2.3$^{\circ}$/0.6$^{\circ}$ & 0.8$^{\circ}$/0.4$^{\circ}$ & 1.0$^{\circ}$/0.4$^{\circ}$ & 0.7$^{\circ}$/0.4$^{\circ}$ \\
\hline
SGCQ9 HF & 1.7$^{\circ}$/0.6$^{\circ}$ & 0.8$^{\circ}$/0.4$^{\circ}$ & 0.6$^{\circ}$/0.4$^{\circ}$ & 0.6$^{\circ}$/0.4$^{\circ}$ \\
\hline
\end{tabular}
\end{center}
\caption{\small{\textit{Pointing accuracy  (A+B+C / A) for circular search.}}}
\label{Tab:circular}
\end{table}

\subsection{Waveform Reconstruction}\label{WFRec} 

\begin{figure}[!hbt]
 \begin{center}
 \includegraphics[width=0.5\textwidth]{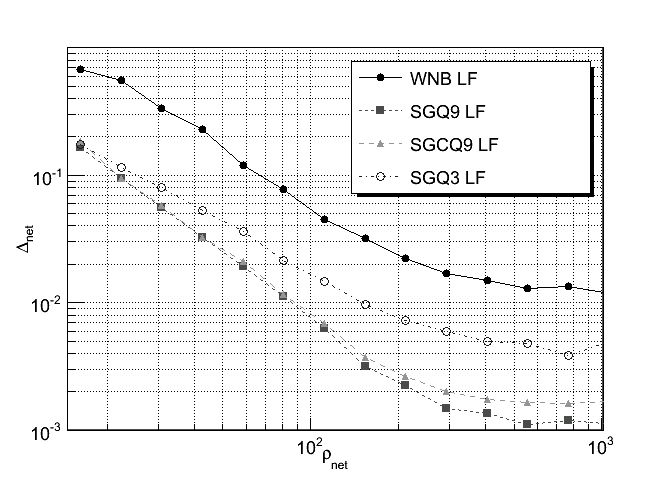}
 \end{center}\caption{\small{\textit{Normalized residual energy $\Delta$ vs network SNR obtained with 
the unmodeled algorithm and the LHV network for all types of injections}}}
 \label{WFr_LHV_LF}
\end{figure}

The signal waveforms are obtained from the solution of the likelihood functional.
Those are the reconstructed detector response as defined by Eq.~\ref{eq:xi}.
To characterize the algorithm performances on waveform reconstruction we consider 
the normalized residual energy $\Delta$:
\begin{equation}
\label{eq:Norm}
\Delta=\frac{({\bf{\xi_{\text{h}}}}-{\bf{\xi}}|{\bf{\xi_{\text{h}}}}-{\bf{\xi}})}
{({\bf{\xi_{\text{h}}}}|{\bf{\xi_{\text{h}}}})} \;,
\end{equation}
where the inner products are defined by Equation~\ref{eq:inner}.
The same as for the coordinate reconstruction, the accuracy of the waveform reconstruction
strongly depends on the strength of detected signal and the waveform morphology. 
At low SNR, the reconstruction is affected by the detector noise, however it improves 
with the increasing value of SNR (see Figure~\ref{WFr_LHV_LF}). At high SNR, the reconstruction
accuracy reaches a limit due to the finite precision of the algorithm. 
In this paper we do not present a detailed study of the waveform reconstruction.
However, such work is planned in the future.

\section{Factors limiting reconstruction} \label{limitation}

In this section we describe the factors which limit the accuracy of the coordinate 
reconstruction including: a) angular and strain sensitivity of the detectors, b) polarization content
of the signals, c) calibration uncertainties and d) limitations of the reconstruction algorithm.

\subsection{Antenna patterns and detector noise}
The angular and the strain sensitivity 
of the network is fully characterized by its noise-scaled
antenna pattern vectors (see Eq.~\ref{eq:apn}). Due to unfortunate sky location or elevated
detector noise, the components of these vectors corresponding to an individual detector may be
small with respect to the other detectors, effectively excluding this detector
from the detection and reconstruction of a marginal GW signal.  For example, 
for a source at ($\theta=-40^o,\phi=50^o$) the angular sensitivity of the V1 detector is 
too low ($\sqrt{|F_+|^2+|F_\times|^2} \sim 0$) and it can not contribute to the reconstruction 
unless the GW signal is very strong. For the LHV network it means that 
for a significant fraction of the sky the triangulation is performed with only two detectors,
which is not sufficient for the accurate source localization. For this reasons it
is very desirable to have four or more detectors in the network 
operating in coincidence.

\subsection{Waveforms and polarization}
\label{LF_WandP}
\begin{figure}[!hbt]
 \begin{center} 
  \begin{tabular}{c}
 \includegraphics[width=80mm]{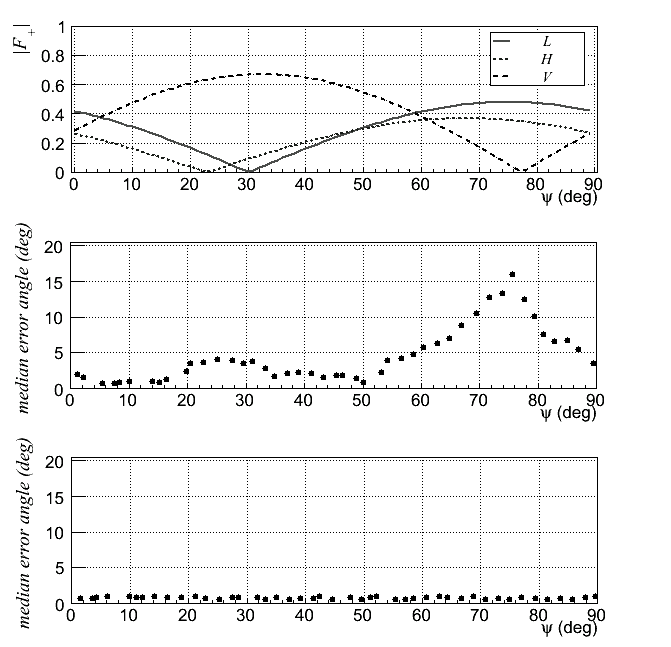}
   \end{tabular}
 \end{center}\caption{\small{\textit{Dependence of $F_+$ 
(top plot, at $\theta=-30^{\circ}$ and $\phi=72^{\circ}$) and 
the reconstruction accuracy on the polarization angle $\psi$ for two types 
of waveforms: linearly polarized SGQ9 LF waveforms (middle) and 
randomly polarized WNB LF waveforms (bottom).}}}
\label{Fig. waveform_polarization}
\end{figure}

For a given direction to the source, the reconstruction accuracy may be very different
depending on the signal polarization. For example, a linearly polarized signal ($h_+,0)$ may not be 
measured by one of the detectors for some values of the polarization angle when 
$|F_{+}|$ is almost null. 
As a result, with the 3-site networks the source localization for such signals
is expected to be poor (see Figure~\ref{Fig. waveform_polarization}).  
On contrary, signals with two polarization components 
can be localized well via their cross component even at the minimum of $F_+$.

\subsection{Calibration uncertainties}
The coordinate reconstruction may be affected by the calibration uncertainties of recorded data streams.
Typically the calibration uncertainties are of the order of 10\% in the amplitude 
and few degrees in the phase~\cite{S5-calib}. 
These systematic distortions of the GW signal may result in a systematic shift of the 
reconstructed sky location away from a true source location.

To estimate this effect we introduce a 
variation of the amplitude and phase of the injected detector responses.
The amplitude variation is selected randomly between $\pm 10\%$ or $0\%$
and the random time shifts of $\pm 32~\mu{s}$ or 0~$\mu{s}$ are introduced. 
The non-zero time shifts correspond to a phase shift of 
$\pm 2.5^{\circ}$ and $\pm 11.5^{\circ}$ at the low and high frequencies respectively. 
Such ``misscalibration'' is applied to all detectors in the network. 
The results are reported in table~\ref{CalErrs} for the LHV network. They show that
depending on the signal morphology and bandwidth the calibration uncertainties 
may affect the coordinate reconstruction.
The effect of calibration uncertainties is particularly visible at high SNR 
where the angular resolution is less affected by the detector noise.  
Similar studies for the LHVAJ network do not show, even at high SNR, any significant
impact of the calibration errors used in the analysis.
This is an expected result, because a better constrained
LHVAJ network provides more robust source localization than the LHV network.

\begin{table}[|h!btc]
\begin{center}
\begin{tabular}{|c|c|c|c|c|}
\hline
Waveform & -  & Amplitude & Phase\\
\hline
WNB LF  &  0.7$^{\circ}$ & 1.0$^{\circ}$ & 0.9$^{\circ}$ \\
\hline
WNB HF  &  0.4$^{\circ}$ & 0.6$^{\circ}$ & 0.8$^{\circ}$ \\
\hline
SGQ9 LF &  0.7$^{\circ}$ & 2.8$^{\circ}$ & 1.2$^{\circ}$ \\
\hline
SGQ9 HF &  0.9$^{\circ}$ & 1.6$^{\circ}$ & 1.4$^{\circ}$ \\
\hline
SGQ3 LF &  0.5$^{\circ}$ & 2.1$^{\circ}$ & 1.0$^{\circ}$ \\
\hline
SGQ3 HF &  0.4$^{\circ}$ & 1.1$^{\circ}$ & 1.0$^{\circ}$ \\
\hline
SGCQ9 LF & 0.8$^{\circ}$ & 2.5$^{\circ}$ & 1.1$^{\circ}$ \\
\hline
SGCQ9 HF & 1.1$^{\circ}$ & 1.9$^{\circ}$ & 2.0$^{\circ}$ \\
\hline
\end{tabular}
\end{center}
\caption{\small{\textit{Pointing accuracy (fit parameter A) for the LHV network and
different signals (column 1): no calibration errors (column 2), 
amplitude and phase mis-calibration (columns 3 and 4 respectively).}}}
\label{CalErrs}
\end{table}

\subsection{Reconstruction algorithm}
There are several factors limiting the accuracy of the coordinate and waveform reconstruction
due to the cWB algorithm. 
For high SNR events the coordinate resolution is limited by
the cWB sky segmentation which is $0.4 \times 0.4$ degrees. Therefore the error angle can 
not be less than $0.4$ degrees. Also for the high frequency events the coordinate resolution 
is limited by the discrete time delays $\tau_k$ (see section~\ref{sec:algorithm}) with the
step of 1/16384 seconds and by the accuracy of the time delay filter (few percent) used in the analysis.
Also in the analysis we did not use any unmodeled constraint specific for individual networks, 
which, in principle, may improve reconstruction.
These limitations are not fundamental and the algorithm performance
can be improved in the future. 
\section{Conclusion}
\label{conclusions}

In the paper we present the results of the source localization and reconstruction of GW
waveforms with the networks of GW interferometers. For a general characterization of 
the detector networks we introduce few fundamental network parameters, including 
the effective noise, and the network antenna and alignment factors. The effective power 
spectral density of the network noise determines the average network SNR for a given 
population of GW signals. For each direction in the sky the network performance 
is characterized by its antenna and alignment factors.
The antenna factor describes how uniform is the network response across the sky.
The alignment factor, which strongly depends on the number of detectors
and the orientation of their arms, determines the relative contribution
of the two GW polarizations into the total network SNR. 

It requires several  non-aligned detectors 
(preferably more than three) for a robust detection and reconstruction of both GW components.  
The coordinate reconstruction strongly depends on the signal waveforms, 
network SNR and the number of detector sites in the network. The reconstruction can be significantly 
improved when it is constrained by the signal model.
Although a crude coordinate reconstruction (ring in the sky) is possible with the networks of two 
spatially separated sites, at least three detector sites are required to perform the source
localization. The accuracy of the localization dramatically increases for networks with more 
than three sites, particularly for the low SNR events. For example, the LH\~{H}V and LHVA
networks are expected to have about the same detection rates, however, the 4-site LHVA
network would have much better performance for the accurate reconstruction of GW signals.  
The pointing resolution required for joint observations with the electromagnetic telescopes 
is achievable with the networks consisting of four sites. 
The LHVAJ network demonstrates further improvements,
both in the detection and reconstruction of GW signals, reaching a sub-degree angular
resolution. In addition, due to the limited
duty cycle of the detectors, both the LCGT and the Australian detectors will 
significantly increase the observation time when any of 4-site networks are operational. 

The advanced LIGO and Virgo detectors are very capable of the first
direct detection of gravitational waves. However, for better reconstruction of
the GW signals more detectors are required. Extra detectors introduce an important 
redundancy which lower the impact of limited duty cycle of the detectors,
makes the coordinate reconstruction more accurate, and less dependent 
on the waveform morphology and calibration uncertainties.
The construction of the LCGT and the detector in Australia 
will significantly enhance the advanced LIGO-Virgo network and these detectors will play
a vital role in the future GW astronomy.

\section{Acknowledgements}
The authors are thankful to the LIGO-Australia committee members R.~Weiss, P.~Saulson,
S.~Sathyaprakash, F.~Raab, P.~Fritschel and S.~Finn for useful discussion of the results. 
Also the authors appreciate suggestions by L.~Bildsten
on better presentation of the results, particularly in Figure~\ref{LarsHistograms}.
This work was supported by the US National Science Foundation grants PHY-0855044 and
PHY-0855313 to the University of Florida, Gainesville, Florida.

\end{document}